\documentclass[%
 reprint,
 amsmath,amssymb,
 aps,
]{revtex4-2}

\usepackage{graphicx}% Include figure files
\usepackage{dcolumn}% Align table columns on decimal point
\usepackage{bm}% bold math
\usepackage[usenames,dvipsnames]{xcolor}
\usepackage{url}
\usepackage{hyperref}
\usepackage{lipsum}

\usepackage{empheq} % for braces and labels

\usepackage{float}

\usepackage{comment}

\usepackage{amsthm}

\begin{document}

\preprint{APS/123-QED}

\title{Static and spherically symmetric vacuum spacetimes with non-expanding principal null directions in $f(R)$ gravity}% Force line breaks with \\

\author{Alberto Guilabert}
\altaffiliation{email: alberto.guilabert@ua.es}
\address{Fundacion Humanismo y Ciencia, Guzmán el Bueno, 66, 28015 Madrid, Spain}
\address{Departamento de F\'{\i}sica Aplicada, Universidad de Alicante, Campus de San Vicente del Raspeig, E-03690 Alicante, Spain}

\author{Pelayo V. Calzada}
\altaffiliation{email: pelayo.delvalle@ua.es}
\address{Departamento de F\'{\i}sica Aplicada, Universidad de Alicante, Campus de San Vicente del Raspeig, E-03690 Alicante, Spain}

\author{Pedro Bargueño}
\altaffiliation{email: pedro.bargueno@ua.es}
\address{Departamento de F\'{\i}sica Aplicada, Universidad de Alicante, Campus de San Vicente del Raspeig, E-03690 Alicante, Spain}

\author{Salvador Miret-Artés}
\altaffiliation{email: s.miret@iff.csic.es}
\address{Instituto de Física Fundamental, Consejo Superior de Investigaciones Científicas, Serrano 123, 28006, Madrid, Spain}

\date{\today}% It is always \today, today,

\begin{abstract}
In this work we characterize all the static and spherically symmetric vacuum solutions in $f(R)$ gravity when the principal null directions of the Weyl tensor are non-expanding. In contrast to General Relativity, we show that the Nariai spacetime is not the only solution of this type when general $f(R)$ theories are considered. In particular, we find four different solutions for the non-constant Ricci scalar case, all of them corresponding to the same theory, given by $f(R) = r_0^{-1}\left\lvert R-3/r_0^2\right\rvert^{1/2}$, where $r_0$ is a non-null constant. Finally, we briefly present some geometric properties of these solutions.
\end{abstract}

\maketitle

\section{Introduction}\label{sec:introduction}

The Nariai spacetime, presented back in the 1950's by Nariai \cite{Nariai1950,Nariai1951}, can be described in suitable coordinates by the line element
\begin{equation}\label{nariai_metric}
ds^2 = \left(1-\dfrac{r^2}{r_0^2}\right) dt^2 - \left(1-\dfrac{r^2}{r_0^2}\right)^{-1} dr^2 - r_0^2 d\Omega^2,
\end{equation}
with $r\in(-r_0,r_0)$, where $r_0$ is a non-null constant, and $d\Omega^2$ is the line element of the 2-sphere. This spacetime is a well known static and spherically symmetric solution for General Relativity (GR) with positive cosmological constant $\lambda=1/r_0^2$.

This spacetime is usually characterized as a special limit of the Schwarzschild–de Sitter solution when the event and cosmological horizons coincide, see e.g. \cite{perry}. Nevertheless, this limit is not defined in a meaningful sense, as it has been pointed out in \cite{2Horizons}. In fact, the resulting spacetimes depend on the coordinate choice when taking limits in the metric tensor \cite{Geroch1969}.

Therefore, we may turn to Petrov classification and the Newman-Penrose (NP) formalism as elegant tools to fully characterize this solution, given their usefulness in the study of algebraically special spacetimes.

Considering the Weyl decomposition of the Riemann tensor, Petrov developed a classification of spacetimes \cite{Petrov} examining the algebraic structure of the curvature tensor. This classification can be carried out by studying the eigenbivectors of the Weyl tensor, which are associated to four null vectors that determine the so called principal null directions of the Weyl tensor. When there is at least one degeneracy between the four principal null directions the spacetime is said to be algebraically special. Moreover, the Petrov classification was also studied later in terms of spinors, see e.g. \cite{Witten,Penrose}, and recently extended to higher dimensions \cite{WeylHigherDim}.

Every spherically symmetric spacetime is algebraically special \cite{Wald}, which naturally induces the use of the NP formalism. In particular, static and spherically symmetric spacetimes are type D (or O). That is, there are two degenerated principal null directions (or the spacetime is conformally flat, which corresponds to type O). As it is usual in the NP formalism, we construct a null tetrad with two null vectors $l^{\mu}$ and $n^{\mu}$, which we take to be aligned with the two principal null directions \cite{NewmanPenrose}. The tetrad is completed by combining a pair of real orthogonal spacelike unit vectors to define a complex null vector $m^{\mu}$ and its complex conjugate $\bar{m}^{\mu}$. Such a null tetrad allows us to take advantage of the symmetries of the Weyl tensor in algebracially special solutions.

In the spherically symmetric case, the congruences associated to the principal null directions are indeed geodesic, non-rotating and shear-free.  In terms of the optical scalars in the NP formalism that is $\kappa = \omega = \sigma = 0$.  In particular, for static and spherically symmetric spacetimes, $\rho$ will determine the congruence expansion of $l$ and $n$ \cite{GHPBook}. In this context, we restrict our analysis to spacetimes whose principal null directions are non-expanding. That is a subset of Kundt class spacetimes, which are defined as the ones admitting a non-expanding, non-rotating and shear-free congruence of null geodesics, see e.g. \cite{KramersExactSolutions}. With the previous assumptions, $\rho=0$ implies that the spacetime can be decomposed as the direct product of two 2-spaces.

The Nariai spacetime, introduced in Eq. \eqref{nariai_metric}, is the paradigmatic prototype for a solution with the previous properties in GR. In fact, it is the only static and spherically symmetric solution with non-expanding principal null directions in GR with cosmological constant \cite{Birkhoff}. Moreover, its uniqueness has also been recently proved in larger dimensions \cite{UniqLargerDim}.

At this point, the following question naturally arises: is the Nariai solution the only static and spherically symmetric vacuum spacetime with non-expanding principal null directions beyond GR?

In this context, Extended Theories of Gravity (ETGs) have been developed to approach some known problems associated to GR, such as the existence of black hole and cosmological singularities \cite{HawkingEllis} or the lack of a satisfactory description for the accelerated expansion of the universe. Among different ETGs (see e.g. the standard reference from Capozziello and Faraoni \cite{CapozzielloBEG} and references therein), two paradigmatic examples are scalar-tensor \cite{ExtendedTheories} or $f(R)$ theories \cite{fRreview}.

In particular, $f(R)$ theories can be useful as toy models which allow us to study gravity modifications in a simplified way. Even more, it is known that the only possibility to obtain a potentially stable local modification of GR is to make the Lagrangian an arbitrary function of the Ricci scalar \cite{Woodard}.

Through the last years, many exact solutions of $f(R)$ gravity have been discovered \cite{CapozzielloBEG,fRreview,ExactSolsfR}. For static and spherically symmetric vacuum spacetimes, the case with expanding principal null directions have been studied by Multamäki and Vilja \cite{Multamaki} and by Sebastiani and Zerbini \cite{sssSebastiani}. The non-vacuum case has also been studied describing the equilibrium configuration of a star \cite{Multamaki_fluid,sssSolutionsfR_star}.

Regarding the vacuum case, it has been shown that constant Ricci scalar solutions are indeed Einstein spaces, which is the case of the Nariai spacetime. In this case, an existence condition for compatible $f(R)$ theories has been given \cite{BarrowNariai,ZerbiniOneLoop,SebastianiInstabilities}. 

Following this motivation, our aim is to fully characterize all the static and spherically symmetric vacuum solutions with non-expanding principal null directions for $f(R)$ theories. 

This manuscript is organized as follows: In Sec. \ref{sec:fRgravity} we give a brief introduction to the $f(R)$ gravity formalism and the corresponding field equations. In Sec. \ref{sec:fieldeq} we construct the line element for a static and spherically symmetric spacetime when the non-expanding condition on the principal null directions is considered. Then, we present the field equations in the new basis constructed with the principal null directions and we solve for the metric, giving the compatible $f(R)$ theories. In particular, we briefly comment some relevant geometric properties about the solutions. Finally, in Sec. \ref{discussion} we point out the main results.

Through all the manuscript we use the signature (1,3) for the metric and the Penrose and Rindler's sign convention for the Riemann curvature tensor \cite{PenroseSpinors}. 

\section{$f(R)$ gravity}\label{sec:fRgravity}

For the vacuum case, the general action in $f(R)$ formalism is \cite{CapozzielloBEG}
\begin{equation}
S[g_{\mu \nu}] = \int d^4x \sqrt{-g}f(R),
\end{equation}
where we have set $16 \pi G=1$, with $G$ denoting the gravitational constant.

\medskip
Applying the variational principle $\delta S=0$, the field equations for $f(R)$ gravity are
\begin{equation}\label{InitialFieldEqs}
     -R_{\mu \nu} = \Delta t_{\mu \nu},
\end{equation}
where we have defined the effective stress-energy tensor as
\begin{multline}\label{stress_energy}
     \Delta t_{\mu \nu} \equiv F(R)^{-1} \Big( -\frac{1}{2} f(R) g_{\mu \nu}\\ +\big[\nabla_{\mu} \nabla_{\nu}-g_{\mu \nu} \square\big] F(R) \Big),
\end{multline}
with $F(R) \coloneqq df(R)/dR$. Then, the associated trace equation is
\begin{equation}\label{trace_eq}
R = F(R)^{-1}\left(2 f(R) + 3\square F(R)\right).
\end{equation}

In the following section we rewrite these field equations employing the NP formalism for our particular case of study.

\section{static and spherically symmetric vacuum solutions with non-expanding principal null directions}\label{sec:fieldeq}

We start with a general static and spherically symmetric spacetime $(M,g)$, whose line element is given by
\begin{equation}\label{metric:staticspherical}
ds^2 = p(r)dt^2-s(r)dr^2-q(r)d\Omega^2,
\end{equation}
where $p(r), s(r)$ and $q(r)$ are positive functions.

In this case, the principal null directions are
\begin{eqnarray}\label{initialbasis}
l_{\mu} & = & \dfrac{1}{\sqrt{2}}\left(p(r)^{1/2},s(r)^{1/2},0,0\right),\nonumber\\
n_{\mu} & = & \dfrac{1}{\sqrt{2}}\left(p(r)^{1/2},-s(r)^{1/2},0,0\right),
\end{eqnarray}
and we complete the null tetrad with 
\begin{equation}\label{orthvect}
m_{\mu} = \dfrac{q(r)}{\sqrt{2}}\left(0,0,1,-i\sin\theta\right),
\end{equation}
where $l_{\mu}n^{\mu} = -m_{\mu}\bar{m}^{\mu} = 1$.

Imposing the non-expanding condition on $l$, \textit{\textit{i.e.}} $\rho=0$, it is straightforward to verify that $q(r)=r_0^2$, where $r_0$ can be taken as a positive constant without loss of generality. Note that this choice of $q(r)$ assures the non-expanding property of $n$.

This case cannot be transformed by a redefinition of coordinates into the expanding case with $q(r) = r^2$, which is usually considered in spherical symmetry, as the aforementioned expansion, $\rho$, is independent of the coordinate choice. Note that this distinction was discussed in detail in \cite{HawkingEllis,KramersExactSolutions,EditorsNote} and more recently in \cite{Birkhoff}. In addition, this differentiation was explicitly studied by Kinnersley when solving for all Type D vacuum metrics within GR \cite{Kinnersley}. Unfortunately, the ansatz with $q(r)=r^2$ is still generally assumed as the most general static and spherically symmetric spacetime, despite the fact that these two cases are not related.

After considering the non-expanding condition, the resulting spacetime $(M,g)$ can be decomposed as a product manifold $M = N\times S^2(r_0)$ with $g=g_N\oplus g_{S^2(r_0)}$, where $g_N$ is the metric of a Lorentzian 2-surface and $g_{S^2(r_0)}$ is the metric tensor of the Riemannian 2-sphere of radius $r_0$.

Therefore, the change of coordinates
\begin{equation}
\Tilde{r}(r) = \int_{\alpha_0}^r du \left(s(u)p(u)\right)^{1/2}
\end{equation}
transforms the metric given in Eq. \eqref{metric:staticspherical} into
\begin{equation}\label{Isothermal_metric}
ds^2 = p(r)dt^2 - p(r)^{-1}dr^2 - r_0^2 d\Omega^2,
\end{equation}
after renaming $\Tilde{r}$ as $r$ \footnote{The function $\Tilde{r}(r)$ is strictly increasing so there exists its inverse function $r(\Tilde{r})$.}. 

As commented in Sec. \ref{sec:introduction}, we present the field equations given in Eq. \eqref{InitialFieldEqs} in the null tetrad basis defined in Eqs. \eqref{initialbasis} and \eqref{orthvect}. Note that in these new coordinates this basis is expressed as
\begin{eqnarray}\label{nullbasis}
l_{\mu} & = & \dfrac{1}{\sqrt{2}}\left(p(r)^{1/2},p(r)^{-1/2},0,0\right),\nonumber\\
n_{\mu} & = & \dfrac{1}{\sqrt{2}}\left(p(r)^{1/2},-p(r)^{-1/2},0,0\right),\\
m_{\mu} & = & \dfrac{r_0}{\sqrt{2}}\left(0,0,1,-i\sin\theta\right).\nonumber
\end{eqnarray}

In order to state the field equations, we introduce the following notation.

On one hand, the so called Ricci scalars, $\Phi_{ab}$ with $a,b\in\{0,1,2\}$, are defined as the contractions of the Ricci tensor with the null tetrad vectors (see e.g. \cite{NewmanPenrose, GHPBook}). These are the components of the Ricci tensor in the new basis introduced in Eq. \eqref{nullbasis}. For the metric given by Eq. \eqref{Isothermal_metric} the only non-vanishing scalar is
\begin{equation}
\Phi_{11} = -\dfrac{1}{2}R_{\mu\nu}l^{\mu}n^{\nu} + 3\Lambda,
\end{equation}
where $\Lambda = R/24$, with $R$ being the Ricci scalar curvature.

On the other hand, we define the \textit{physical contractions} of the effective stress-energy tensor in an analogous way as the Ricci scalars. Using the field equations given in Eq. \eqref{InitialFieldEqs}, the only non-vanishing scalars are
\begin{eqnarray}
\Phi^{ph}_{00} & = & \dfrac{1}{2}\Delta t_{\mu\nu}l^{\mu}l^{\mu},\\
\Phi^{ph}_{11} & = & \dfrac{1}{2}\Delta t_{\mu\nu}l^{\mu}n^{\nu} + 3\Lambda^{ph},\\
\Phi^{ph}_{22} & = & \dfrac{1}{2}\Delta t_{\mu\nu}n^{\mu}n^{\nu},
\end{eqnarray}
with $\Lambda^{ph}=R/24$, where $R$ is now obtained in terms of $f(R)$ by using Eq. \eqref{trace_eq}.

At this point, the only field equations which are not identically zero in the new basis are given by $\Phi^{ph}_{00} = \Phi^{ph}_{22} = 0$, $\Phi^{ph}_{11} = \Phi_{11}$ and $\Lambda^{ph} = \Lambda$. For convenience we will take the following linear combinations:

\begin{itemize}
    \item $\Phi^{ph}_{00} = \Phi_{00}$
    \begin{equation}\label{v2_eq:phi00}
        F''(r) = 0,
    \end{equation}
    \item $\Phi^{ph}_{11} = \Phi_{11}$
    \begin{equation}\label{v2_eq:phi11}
        -\dfrac{F'(r)p'(r)+p(r)F''(r)}{F(r)} = \dfrac{2}{r_0^2}+p''(r),
    \end{equation}
    \item $\Phi^{ph}_{11}-\Lambda^{ph} = \Phi_{11}-\Lambda$
    \begin{equation}\label{v2_eq:phi11-l}
        -\dfrac{f(r)}{2F(r)} = \dfrac{1}{r_0^2}+p''(r),
    \end{equation}
\end{itemize}
where $F(r) \equiv F(R(r))$ and $f(r) \equiv f(R(r))$.

Observe that Eq. \eqref{v2_eq:phi00} implies
\begin{equation}\label{F(r)teoria}
F(r) = a(1+br),
\end{equation}
where $a$ and $b$ are constants having dimensions of one over length.

In addition, from Eq. \eqref{F(r)teoria} it is natural to consider two different cases depending whether $b=0$ or $b \neq 0$. Interestingly, as we show along this section, $b=0$ corresponds to constant Ricci scalar and  $b\neq 0$ to non-constant Ricci scalar. This will allow us to fully characterize the solutions in terms of the Ricci scalar. 

\subsection{Non-constant Ricci scalar solutions}

Assuming $b\neq 0$ in Eq. \eqref{v2_eq:phi11}, the solution is given by
\begin{equation}\label{p(r)solution}
p(r) = c_1 - \dfrac{r}{b r_0^2} - \dfrac{r^2}{2r_0^2} + \dfrac{\gamma}{b^2 r_0^2} \log\left|1+br\right|.
\end{equation}
where $\gamma = 1+c_2 b r_0^2$, being $c_1$ and $c_2$ two arbitrary constants. Indeed, their relevance on the properties of the resulting spacetimes is shown below.

Observe that in the limit case $b\to 0$ on Eq. \eqref{p(r)solution} the Nariai solution is recovered as
\begin{equation}\label{limit_case_p(r)}
\lim_{b\to 0} p(r) = c_1 - c_2 r - \dfrac{r^2}{r_0^2},
\end{equation}
which always can be transformed into Eq. \eqref{nariai_metric} by a suitable change of coordinates.

Moreover, the Ricci scalar is given by
\begin{equation}\label{ricciscalarnonconstant}
R(r) = \dfrac{1}{r_0^2}\left(3 + \dfrac{\gamma}{\left(1 + b r\right)^2}\right),
\end{equation}
and it shows a curvature singularity at $r = -1/b$.

Now we focus on the study of the compatible $f(R)$ theories. Eq. \eqref{v2_eq:phi11-l} can be written in terms of the Ricci scalar as
\begin{equation}\label{DifEq:fR}
R-\dfrac{3}{r_0^2} = \dfrac{f(R)}{2f'(R)},
\end{equation}
and from this equation we deduce 
\begin{equation}\label{fR_theory}
f(R) = \dfrac{\alpha}{r_0}\left\lvert R-\dfrac{3}{r_0^2} \right\rvert^{1/2},
\end{equation}
where $\alpha$ is a dimensionless non-null constant. We take $\alpha = 1$ without loss of generality \footnote{A constant multiple of the action does not modify the field equations when the variational principle is applied.}. Also note that GR is not recovered in the limit case when $b\to 0$ shown in Eq. \eqref{limit_case_p(r)} since the theory, determined by Eq. \eqref{fR_theory}, is independent of $b$.

It remains to verify the consistency between Eqs. \eqref{F(r)teoria}, \eqref{ricciscalarnonconstant} and \eqref{fR_theory}. That is, 
\begin{equation}
\frac{df}{dR}(R(r)) = F(r),
\end{equation}
which yields
\begin{equation}
F(r) = \dfrac{\gamma}{2 \left|\gamma\right|^{3/2}} \left|1+br\right|,
\end{equation}
from where $a = \text{sign}(1+br)\gamma\left|\gamma\right|^{-3/2}/2$. Note that $a$ is constant in each connected subinterval of the metric domain (as $r=-1/b$ is not included). Moreover, $a\in\mathbb{R}$ imposes $\gamma \neq 0$. Therefore, by Eq. \eqref{ricciscalarnonconstant}, the Ricci scalar is non-constant.

The solution given in Eq. \eqref{p(r)solution} has, if any, the following Killing horizons generated by $\partial_t$, which are determined by
\begin{equation}\label{Killing_Horizons}
r_h = -\dfrac{1}{b}\left(1\pm\sqrt{-\gamma W_{k}\bigl(\xi(c_1,\gamma)\bigr)}\right),
\end{equation}
where we have defined
\begin{equation}\label{eq:xi}
\xi(c_1,\gamma) = -\dfrac{1}{\gamma}e^{\frac{-1-2 c_1 b^2 r_0^2}{\gamma}}
\end{equation}
and $W_k$ is $k$-th branch of the Lambert's function. As $r$ must be a real value, the only considered branches are $k \in \{-1,0\}$.

\begin{figure}
    \centering
    \includegraphics[width=0.45\textwidth]{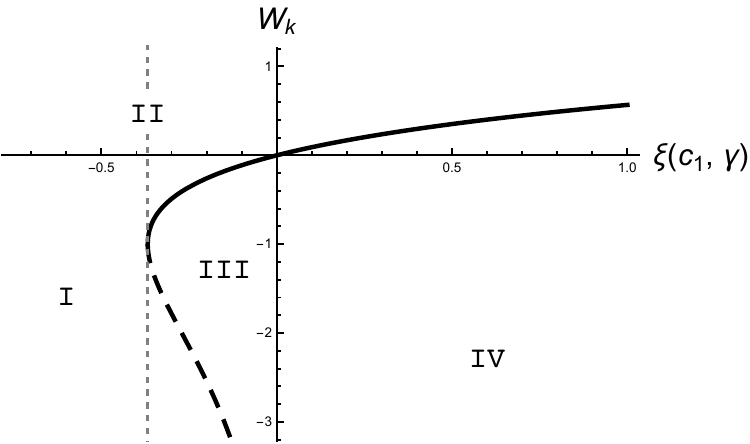}
    \caption{Lambert's $W$ function for the two branches $k\in\{0,1\}$. The principal and $k=-1$ branches are represented in solid and dashed lines, respectively. Case $\texttt{I}$ is defined when no real value of the Lambert function exists. In case $\texttt{II}$ the real value of the two branches coincides. Case $\texttt{III}$ determines two different real values for the Lambert $W$ function. Finally, case $\texttt{IV}$ is defined when the principal branch is the only real value. Note that, in accordance with Eq. \eqref{eq:xi}, $\xi(c_1,\gamma) = 0$ is excluded.}
    \label{fig:regions}
\end{figure}

At this point, we differentiate into four different cases for $\xi(c_1,\gamma)$. Each case represents the number of real values of the Lambert $W$ function for the two considered branches, as Fig. \ref{fig:regions} shows. This determines the number of Killing horizons, according to Eq. \eqref{Killing_Horizons}.

\begin{itemize}
    \item Case $\texttt{I}$ is determined by $\xi(c_1,\gamma)< -1/e$. Under this condition, there are no Killing horizons generated by $\partial_t$ and the metric function is $p(r)<0$ for all $r\in\mathbb{R}\setminus\{-1/b\}$. Then, no static patch is found.
    
    \item Case $\texttt{II}$ is given by the relation $\xi(c_1,\gamma) = -1/e$. In this situation, there are two different Killing horizons (these are degenerated, in the sense that the two branches coincide) and the metric function is $p(r)\leq 0$ for all $r\in\mathbb{R}\setminus\{-1/b\}$. Thus, there is no static patch in this case.

    \item Case $\texttt{III}$ is defined for $-1/e < \xi(c_1,\gamma) < 0$. There exist four different Killing horizons, which lead to two disconnected static patches.
\end{itemize}

\begin{figure}
    \centering
    \includegraphics[width=0.45\textwidth]{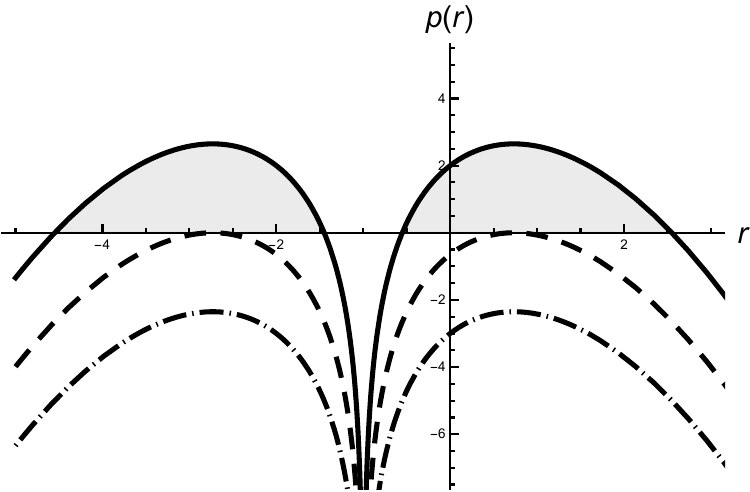}
    \caption{Metric function $p(r)$ against coordinate $r$ obtained for $b = r_0 = 1$, $\gamma = 3$ and different values of $c_1$ in the three possible regions with positive $\gamma$. With solid line we show $c_1$ in region $\texttt{I}$, with dashed line in region $\texttt{II}$ and with dot-dashed line in region $\texttt{III}$. The filled region represents the static patch.}
    \label{fig:p(r)_gamma>0}
\end{figure}

A representative example of the metric function $p(r)$ is shown in Fig. \ref{fig:p(r)_gamma>0} for cases $\texttt{I}$, $\texttt{II}$ and $\texttt{III}$. Observe that in these cases $\gamma >0$, so the Ricci scalar is greater than $3/r_0^2$. Moreover, the curvature singularity lies outside the static region.

\begin{figure}
    \centering
    \includegraphics[width=0.35\textwidth]{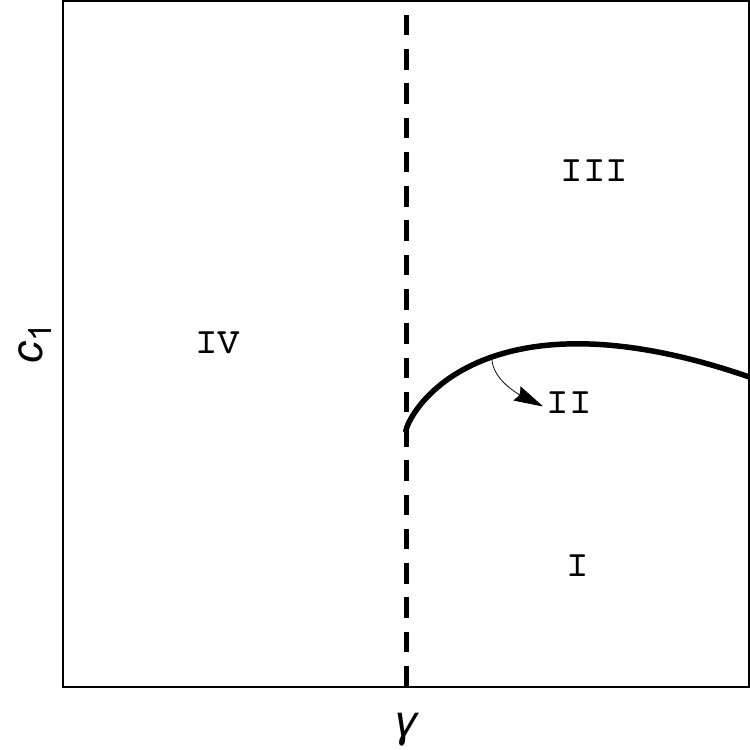}
    \caption{Different values of $c_1$ and $\gamma$ define different cases as shown in Fig. \ref{fig:regions}. For $\gamma > 0$, we can go through cases $\texttt{I}$, $\texttt{II}$ and $\texttt{III}$ by varying $c_1$. Case $\texttt{IV}$ corresponds to $\gamma<0$ regardless of the value of constant $c_1$. Note that it is possible to switch between case $\texttt{IV}$ and the other cases by varying $\gamma$ (or equivalently, $c_2$). The case $\gamma=0$, represented in dashed line, is excluded.}
    \label{fig:regions_c1_gamma}
\end{figure}

Note that by varying $c_1$ it is possible to switch between cases $\texttt{I}, \texttt{II}$ and $\texttt{III}$, as it is shown in Fig. \ref{fig:regions_c1_gamma}. Moreover, as the number of Killing horizons generated by $\partial_t$ must be invariant under local isometries, the constant $c_1$ cannot be removed under changes of coordinates.

\begin{figure}
    \centering
    \includegraphics[width=0.45\textwidth]{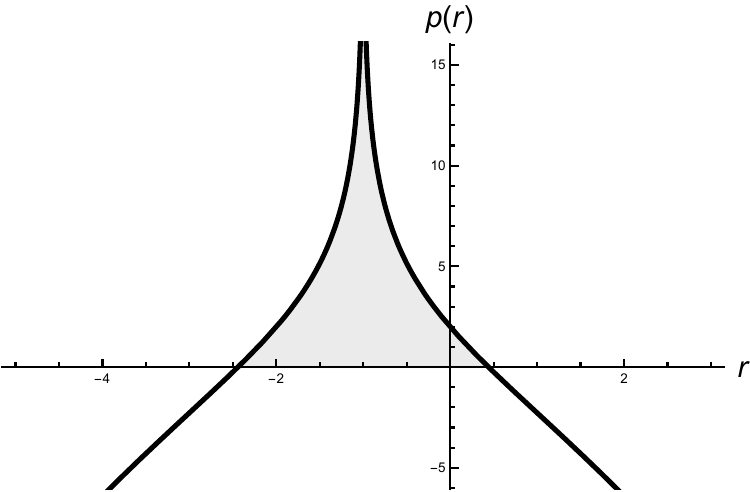}
    \caption{Metric function $p(r)$ against coordinate $r$ in region $\texttt{IV}$ obtained for $b = r_0 = 1$, $c_1 = 2$ and $\gamma = -1$. The filled region represents the static patch.}
    \label{fig:p(r)_gamma<0}
\end{figure}

\begin{itemize}
    \item Case $\texttt{IV}$ is determined by $\xi(c_1,\gamma) > 0$. There are only two different Killing horizons which lead to the existence of one static patch. Moreover, they are independent of the possible values of $c_1$. For illustration, a representation of $p(r)$ is shown in Fig. \ref{fig:p(r)_gamma<0}.
    
    Observe that, by Eq. \eqref{ricciscalarnonconstant}, the Ricci scalar changes its sign at
    \begin{equation}
    r = -\dfrac{1}{b}\left(1\pm\sqrt{-\gamma/3}\right),
    \end{equation}
    and is upper bounded by $3/r_0^2$. Also notice that the constant $c_2$ cannot be removed either by changes of coordinates because by varying $c_2$ it is possible to switch from case $\texttt{IV}$ to the other cases (see Fig. \ref{fig:regions_c1_gamma}) which are not isometric as the different range of the Ricci scalar reveals.
    
    In addition, observe that the curvature singularity at $r=-1/b$ is in this case located inside the static region.
\end{itemize}

Finally, the constant Ricci scalar case will be briefly examined for completeness.

\subsection{Constant Ricci scalar solutions}

Assuming $b=0$, the solution of Eq. \eqref{v2_eq:phi11} is
\begin{equation}
p(r) = c_1 + c_2 r -\dfrac{r^2}{r_0^2},
\end{equation}
where $c_1$ and $c_2$ are arbitrary constants that can be removed by a suitable coordinate changes. Thus, we get
\begin{equation}
p(r) = 1-\dfrac{r^2}{r_0^2},
\end{equation}
which is the Nariai solution introduced in Eq. \eqref{nariai_metric}. Indeed, the Ricci scalar is constant with $R(r) = R_0 \equiv 4/r_0^2$.

From Eq. \eqref{v2_eq:phi11-l} it can be seen that the set of compatible $f(R)$ theories are those functions fulfilling the \textit{one point} differential equation
\begin{equation}\label{fRcondition}
\frac{df}{dR}\left(\frac{4}{r_0^2}\right) = \frac{r_0^2}{2}f\left(\frac{4}{r_0^2}\right).
\end{equation}

This can be rewritten as
\begin{equation}\label{papersConditions}
R_0 \dfrac{df}{dR}(R_0) = 2 f(R_0),
\end{equation}
which is actually the trace equation, introduced in Eq. \eqref{trace_eq}, for constant Ricci scalar. Note that Eq. \eqref{papersConditions} was also presented in \cite{BarrowNariai,ZerbiniOneLoop,SebastianiInstabilities}.

At this point, we can conclude that the only constant Ricci scalar solution to any theory satisfying Eq. \eqref{papersConditions} is the Nariai spacetime. Note that, in general, these theories are not necessarily GR although the corresponding field equations can be interpreted, for constant $R=R_0$, in terms of GR with a cosmological constant given by $\lambda = \frac{R_0}{2} - \frac{f(R_0)}{2F(R_0)}$ \cite{CapozzielloBEG}. In our particular case, Eq. \eqref{papersConditions} implies $\lambda = R_0/4 = 1/r_0^2$, which corresponds to the cosmological constant for the Nariai solution. 

\section{Final Remarks}\label{discussion}

This manuscript aims to extend the study carried out by Multamäki and Vilja \cite{Multamaki} for static and spherically symmetric vacuum spacetimes in $f(R)$ gravity for the case of non-expanding principal null directions, dividing our analysis in two different cases depending on whether the Ricci scalar is constant or not.

For the non-constant Ricci scalar case we have shown that there exists only one $f(R)$ theory which has the form
\begin{equation}
f(R) = \dfrac{1}{r_0}\left\lvert R-\dfrac{3}{r_0^2}\right\rvert^{1/2}.
\end{equation}

This theory admits four different spherically symmetric vacuum solutions (cases $\texttt{I}$--$\texttt{IV}$) with non-expanding principal null directions, which exhibit a curvature singularity at $r=-1/b$. Interestingly, only two of them (cases $\texttt{III}$ and $\texttt{IV}$) present static patches.

In case $\texttt{I}$, no Killing horizons generated by $\partial_t$ are found and there is no static patch. Similarly, no static region is defined in case $\texttt{II}$ despite there exist two Killing horizons. In case $\texttt{III}$, there are four different Killing horizons which define two static patches. Finally, in case $\texttt{IV}$ there are two Killing horizons resulting in a static region.

In cases $\texttt{I}$--$\texttt{III}$, the curvature singularity is found to be outside the static region. On the contrary, in case $\texttt{IV}$, the curvature singularity is located inside the static patch.

We note that our case $\texttt{III}$ with $c_1 = \gamma = 1$ was recently obtained in the context of scale-dependent gravity \cite{Rincon:2023kun}. Given that there is a scale-dependent representation for any $f(R)$ theory \cite{ComparingfR}, this would enable the interpretation of all the solutions obtained in this manuscript in terms of variable gravitational couplings. We left this for future work.

For completeness, we have considered the constant Ricci scalar case, where the Nariai spacetime is shown to be the only static and spherically symmetric vacuum spacetimes with non-expanding principal null directions in $f(R)$ gravity.

Therefore, we can finally assure that the Nariai solution is not the only static and spherically symmetric vacuum spacetime with non-expanding principal null directions in $f(R)$ gravity.

\section*{Acknowledgments}
We acknowledge financial support from the Generalitat Valenciana through PROMETEO PROJECT CIPROM/2022/13. A. G. acknowledges Fundación Humanismo y Ciencia for financial support. P. V. C. acknowledges support from Generalitat Valenciana grant CIACIF/2021/268. A.G. acknowledges M. Carmen and Jant for continuous support.

\bibliography{references}
\end{document}